\documentclass[prl,twocolumn]{revtex4}
\usepackage[dvips]{graphicx}
\usepackage{latexsym}
\usepackage{bm}

\def\etal{~\textit{et~al.}} 
\def\ra{\rangle} 
\def\la{\langle} 

\begin{document}

\title{Exotic order in simple models of bosonic systems}
\author{O. I. Motrunich}
\author{T. Senthil}
\affiliation{Massachusetts Institute of Technology,
77 Massachusetts Ave.,  Cambridge, MA 02139}

\date{May 8, 2002}

\begin{abstract}
We show that simple Bose Hubbard models with unfrustrated hopping
and short range two-body repulsive interactions can support stable
fractionalized phases in two and higher dimensions, and in
zero magnetic field.
The simplicity of the constructed models advances the 
possibility of a controlled experimental realization
and novel applications of such unconventional states.
\end{abstract}

\maketitle

Recent theoretical developments\cite{RSSpN,MoeSon,BalMPAFGir,frcmdl}
have shown that two or three dimensional strongly correlated systems 
in zero magnetic field, could display quantum phases with fractional 
quantum numbers.
This theoretical progress, inspired mostly by the search for a 
theory of the high-temperature superconductors\cite{PWA}, 
is likely to play an important role in our eventual understanding of 
the mysterious properties of several strongly interacting electronic 
systems.
However, to date, no such experimental system has been unambiguously 
shown to display fractional quantum numbers. 
Further impetus for the search for experimental realizations of 
fractionalization comes from the possibility of using such states 
to construct qubits\cite{Kit,Iof}.  
The topological structure inherent in these states naturally protects 
the system from decoherence. 

The primary goal of this paper is the identification and 
possible design of specific condensed matter systems which display 
the phenomenon of fractional quantum numbers. 
To that end, we study particularly simple models of bosons with 
unfrustrated hopping and short ranged two-body repulsive interactions 
on a two-dimensional (2D) square lattice. 
We show that in particular parameter ranges,
a fractionalized insulating phase exists where there are 
excitations whose charge is one half that of the underlying bosons. 
Superfluid or more conventional insulating phases result in other 
parameter ranges.  
The simplicity of our models opens up the possibility that they can be 
realized in arrays of quantum Josephson junctions, or possibly in 
ultra-cold atomic gases. 
This would provide a definite experimental realization of a 
fractionalized phase which could then possibly be exploited to 
construct topologically protected qubits. 

The fractionalized phase appears in a region of intermediate 
correlations where neither the boson kinetic energy nor repulsive 
potential energy completely dominates over the other. 
This lends support to the general notion that fractionalization is 
to be looked for in a many-body system at intermediate correlations.
For example, in the interacting electron system, fractionalization 
possibly occurs at intermediate values of density somewhere between 
the extreme low density Wigner crystal and the high density Fermi 
liquid regimes.  Similarly, electronic Mott insulators that are close 
to the metal-insulator transition may be good candidates for 
fractionalization.

The generalization of our models to three dimensions (3D) is of some 
interest.
The 3D version of our boson Hubbard model has in fact two distinct 
fractionalized insulating phases: 
First, there is a fractionalized phase similar to the one in 2D, 
with the distinct excitations being a charge-$1/2$ chargon and 
a $Z_2$ vortex (vison) line.  The topological order in this phase is 
stable up to a finite non-zero temperature. 
Experimental realization of this phase may therefore be of interest 
for the quantum computing application as a way of controlling errors 
due to non-zero temperature. 
Another {\em distinct} fractionalized insulator also appears in 3D. 
In this phase, the excitations are a gapped charge-$1/2$ chargon, 
a gapless linear dispersing ``photon'', and a gapped topological 
point defect (the ``monopole'').   
Wen\cite{Wen:light} has recently pointed out that stable mean field 
theories may be constructed for quantum phases where a massless 
$U(1)$ gauge boson (a photon) {\em emerges} in the low energy 
description. 
Our results provide an explicit and concrete model for such a phase. 

\begin{figure}
\centerline{\includegraphics[width=2.6in]{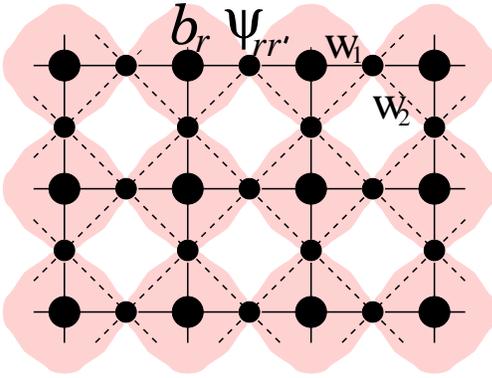}}
\vskip -2mm
\caption{Josephson junction array on a $2$D bond-centered square 
lattice modeled by the Hamiltonian Eq.~(\ref{Hboson}).
Each shaded area indicates schematically cluster charging 
energy $U N_r^2$.
}
\label{lattice}
\end{figure}

\vskip 0.5mm
\textbf{Fractionalization of bosons in two dimensions:}
Consider bosons moving on the lattice shown in Fig.~\ref{lattice}. 
A physical realization may be a Josephson junction array with 
superconducting islands arranged on the sites of the ``bond-centered'' 
square lattice and Josephson-coupled with each other as indicated by 
the links.  We also stipulate repulsive interactions between the 
bosons (``charging energy'') that favors charge neutrality not only 
on individual islands but also on the shaded clusters 
(note that neighboring clusters share one site)\cite{JJK}.
The corresponding Hamiltonian is
\begin{eqnarray}
\label{Hboson}
H & = &
-w_1 \sum_{r, r' \in r} (b_r^\dagger \psi_{rr'} + h.c.)
-w_2 \sum_{[ rr'r'' ]}(\psi_{rr'}^\dagger \psi_{r'r''} + h.c.) 
\nonumber \\
&& + u_b \sum_r (n_r^b)^2 + u_\psi \sum_{\la rr' \ra} (n_{rr'}^\psi)^2 
   + U \sum_r N_r^2 ~.
\end{eqnarray}
Here, $b_r^\dagger = e^{i\theta_r}$ represent bosons (Cooper pairs)
residing on the corner sites of the lattice, 
and $\psi_{rr'}^\dagger=e^{i\phi_{rr'}}$ represent bosons
on the bond-centered sites (identified by the bond end-points);
$n_r^b$, $n_{rr'}^\psi$ are the corresponding boson numbers,
$[\theta_r, n_r^b]=i$, and similarly for the $\psi$-bosons.
Throughout, we work with a number-phase (quantum rotor) representation 
of the bosons, as is particularly appropriate in the 
Josephson junction array realization.

The $w_1$ term is a boson hopping (Josephson coupling) between the 
corner and the bond-centered sites, and $r' \in r$ sums over all 
such bonds emanating from $r$.  
The $w_2$ term is a boson hopping between the neighboring 
bond-centered sites as indicated with dashed lines 
in Fig.~\ref{lattice}.
The $u_b$ and $u_\psi$ terms represent on-site boson repulsion, 
while the $U$ term is the cluster charging energy that favors
charge neutrality in each cluster.
The operator $N_r$ associated with each cluster is defined through 
\begin{equation}
N_r = 2n_r^b + \sum_{r' \in r} n_{rr'}^\psi ~.
\end{equation}
The total boson number of the system is 
$N_{\rm tot} = \frac{1}{2} \sum_r N_r$.
Both the $b$-bosons and the $\psi$-bosons are assigned charge $q_b$.
The model has only a global $U(1)$ charge conservation symmetry.

\begin{figure}
\centerline{\includegraphics[width=2.6in]{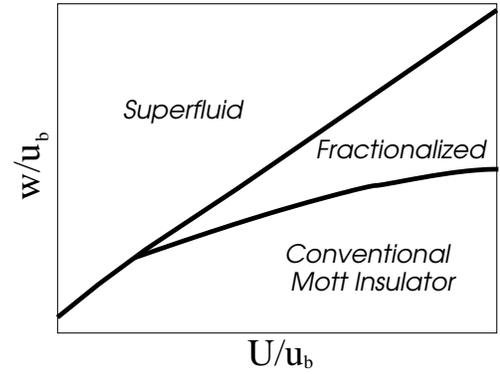}}
\vskip -2mm
\caption{Schematic phase diagram of the boson Hubbard model
Eq.~(\ref{Hboson}) for a particular cut 
$w \equiv w_1 \simeq w_2$ and $u_b \simeq u_\psi$ 
through the parameter space.
}
\label{phased}
\end{figure}

For large $w_1, w_2 \gg u_b, u_\psi, U$ the system is a superfluid.
In the opposite limit, $u_b, u_\psi, U \gg w_1, w_2$, the system is 
a conventional Mott insulator with charge quantized in units of $q_b$.
We argue below that when the charging energies $U$ and $u_b, u_\psi$, 
are varied separately, there is an intermediate regime 
$U \gg w_1, w_2 \gg \sqrt{u_b U}, \sqrt{u_\psi U}$,
in which the system is a stable fractionalized insulator 
with charge $\frac{q_b}{2}$ excitations and charge $0$ visons above 
a ground state with no conventional broken symmetries. 
A schematic phase diagram of our model is shown in 
Fig.~\ref{phased}.

The analysis in the limit of large cluster interaction 
$U \gg w_1, w_2, u_b, u_\psi$ is similar to that in the 
large $U$ limit of the {\em electronic} Hubbard model at half-filling.
If the other terms are all zero, there is a degenerate manifold of 
ground states specified by the requirement $N_r=0$ for each $r$.
This ground state sector is separated by a large charge gap $U$ from 
the nearest sectors.
Including the $w_1, w_2, u_b, u_\psi$ terms lifts the degeneracy 
in each such zeroth-order sector, and this is best described by 
deriving the corresponding effective Hamiltonians for small 
perturbing couplings.  

Consider the ground state sector $N_r = 0$ for all $r$.
An elementary calculation gives 
\begin{eqnarray}
\label{Heff}
H_{\rm eff}^{(0)} = H_{u_b, u_\psi}
-J_{\rm bond} \sum_{\la rr' \ra} 
  \left[ (\psi_{rr'}^\dagger)^2 b_r b_{r'} + h.c. \right]
\nonumber \\
-K_{\rm ring} \sum_\Box 
  \left( \psi_{12}^\dagger \psi_{23} \psi_{34}^\dagger \psi_{41} 
         + h.c. \right) ~,
\end{eqnarray}
where $H_{u_b,u_\psi}$ stands for the on-site repulsion terms
as in Eq.~(\ref{Hboson}),
$J_{\rm bond}=w_1^2/U$, and $K_{\rm ring}=2w_2^2/U$.

\begin{figure}
\includegraphics[width=2.6in]{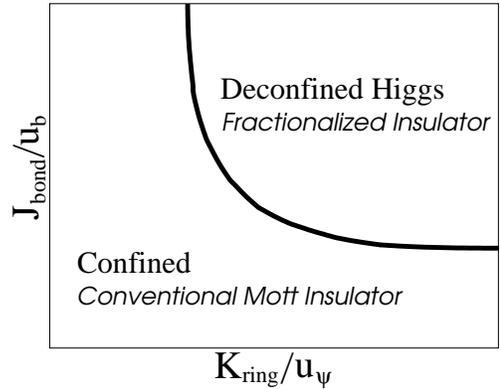}
\vskip -2mm
\caption{Diagram of the large $U$ insulating phases of
the boson model Eq.~(\ref{Hboson}) in two dimensions.
The effective Hamiltonian Eq.~(\ref{Heff}) is equivalent 
to the $(2+1)$D compact QED gauge theory coupled to a charge 2 
scalar.
The Mott insulator is conventional or fractionalized depending on 
whether the effective gauge theory is confined or deconfined.
}
\label{QEDc2_phased_2d}
\end{figure}

A simple change of variables shows\cite{frcmdl} that 
$H_{\rm eff}^{(0)}$ together with the constraint $N_r = 0$ can be 
regarded as the well-studied\cite{FraShe} $(2+1)$D compact $U(1)$ 
gauge theory coupled to a charge 2 scalar field.
In $(2+1)$D, there are two distinct phases shown in 
Fig.~\ref{QEDc2_phased_2d}.
For $J_{\rm bond}, K_{\rm ring} \lesssim u_b, u_\psi$, the
gauge theory is ``confined'', and all excitations carrying
non-zero ``gauge charge''  are confined. Zero gauge charge excitations 
carrying physical charge quantized in units of $q_b$ of course exist 
with a gap of order $2U$.  
This is the conventional Mott insulator of our boson model.

In the opposite regime,
$J_{\rm bond}, K_{\rm ring} \gtrsim  u_b, u_\psi$,
the gauge theory is in the ``deconfined Higgs'' phase.
Objects with $N_r = 1$ at some site, {\em i.e.}~physical charge 
$\frac{q_b}{2}$ (chargon), have gauge charge $1$, are not confined and can propagate above a 
finite gap of order $U$.
There is also a stable gapped $Z_2$ vortex excitation (vison).
The deconfined phase has a topological order\cite{Wen2,topth}: e.g., 
the ground state is two-fold degenerate on a cylinder,
obtained by threading no or one vison through the hole of the 
cylinder.

The details of the chargon motion are determined by the
effective Hamiltonians that obtain in the charged sectors.
Straightforward calculation shows that the presence of the chargon 
induces a weakening of the background on the bonds and plaquettes 
that are connected to it:
$J_{\rm bond}' = \frac{1}{4} J_{\rm bond}$, 
$K_{\rm ring}' = \frac{5}{8} K_{\rm ring}$.
Of course, chargon confinement/deconfinement is controlled entirely 
by the bulk $J_{\rm bond}, K_{\rm ring}$ vs $u_b, u_\psi$ terms that 
obtain far away from the chargon location and is as expected by 
looking at $H_{\rm eff}^{(0)}$ only.  
 
\vskip 0.5mm
\textbf{Ground state wavefunctions and topological order:}
A good caricature for the ground state wavefunction of 
$H_{\rm eff}^{(0)}$, Eq.~(\ref{Heff}), is obtained by 
``Gutzwiller''-projecting a superfluid state into the sector $N_r = 0$
\begin{equation}
\label{GS}
|\Phi \ra \;\;=\;\; P_0 |\theta_r\!=\!\phi_{rr'}\!=\!0 \ra
\;\;= { \sum_{\{n_r^b, n_{rr'}^\psi\}} }\!\!\!\!\!^\prime \;\;
|\{n_r^b, n_{rr'}^\psi \} \ra ~,
\end{equation}
where the last form is written in the boson number basis and the 
primed sum is over all configurations such that $N_r = 0$ at 
every site $r$. 
This is the {\em exact} ground state wavefunction when 
$u_b = u_{\psi}= 0$ but is not normalizable. 
A normalizable wavefunction is obtained by introducing a cutoff for 
large occupation numbers at each site as is appropriate for non-zero 
$u_b, u_{\psi}$.  
Below we leave any such cutoff procedure implicit.

A topologically distinct ground state on a cylinder is obtained by 
Gutzwiller-projecting a superfluid state with one vortex threading 
the cylinder\cite{toGutzw}:
\begin{eqnarray}
|\Phi_v\ra
\;\;= { \sum_{\{n_r^b, n_{rr'}^\psi\}} }\!\!\!\!\!^\prime \;\;
(-1)^{N_{\rm col}^\psi} 
|\{n_r^b, n_{rr'}^\psi \} \ra ~,
\end{eqnarray}
where $N_{\rm col}^\psi$ is a sum of $n_{rr'}^\psi$ in a given 
columnar ``cut'' of $\mathbf{\hat x}$-directed links 
(assuming the cylinder is defined by periodic boundary conditions along 
$\mathbf{\hat x}$). 
Due to the constraints, the parity of $N_{\rm col}^\psi$ is the same 
for all columns so that the location of the cut is arbitrary.  
The projected vortex state describes one vison threading the hole of 
the cylinder.

The presence of topological order is established by
noticing that the normalized overlap 
$\la \Phi | \Phi_v \ra / \la \Phi | \Phi \ra $ 
goes to zero as $O(e^{-cL_y})$ with the system size\cite{unpub},
and that all local physical operators are the same in the two 
states since the column defining $N_{\rm col}^\psi$ can be 
deformed away from any such operator.
Thus, the states with no or one vison are indeed orthogonal to each 
other and degenerate in the thermodynamic limit.

The boson Hamiltonian Eq.~(\ref{Hboson}) is {\em unfrustrated}, 
in the sense that the hopping amplitudes are all positive. 
It is well-known then that the ground state wavefunction is unique 
and has positive amplitudes in the boson number basis.  This does not 
contradict the topological order in the fractionalized state. 
The column parity $(-1)^{N_{\rm col}^\psi}$ is 
conserved by the Hamiltonian $H_{\rm eff}^{(0)}$, and the theorem 
applies to $H_{\rm eff}^{(0)}$ only separately in the even and odd 
sectors on the cylinder.
By taking the combinations $|\Phi\ra \pm |\Phi_v\ra$ we indeed obtain 
positive wavefunctions that reside completely in the even or odd 
sectors.  
As far as the bare boson Hamiltonian Eq.~(\ref{Hboson})
is concerned, it is more appropriate to speak of the states with 
no or one vison.

\vskip 0.5mm
\textbf{Variations in 2D:}
A simple variation of the model considerably enhances the region of 
stability of the fractionalized phase.  Consider ``half-filling'' 
for the site bosons described by the modified Hubbard repulsion terms
\begin{eqnarray}
H_{u_b, U} = u_b \sum_r \left(n_r^b-\frac{1}{2} \right)^2
+ U \sum_r (N_r-1)^2 ~.
\label{Uhfill}
\end{eqnarray}
All other terms are unchanged.  In the large $U$ limit, the 
corresponding compact QED theory now has static charges $\pm 1$ placed 
on the A and B sublattices respectively\cite{frcmdl}, and is at 
half-filling for the gauge charge 2 matter field.
\begin{figure}[b]
\vskip -3mm
\includegraphics[width=2.6in]{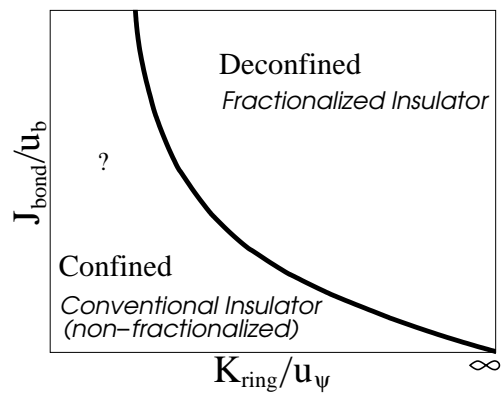}
\vskip -3mm
\caption{Large $U$ insulating phases of the 2D boson model
at half-filling, Eq.~(\ref{Uhfill}).  The region of stability
of the fractionalized phase is enhanced compared with 
Fig.~\ref{QEDc2_phased_2d} at integer filling.
There may be several (?) non-fractionalized insulating states
with broken translational invariance; 
here, we focus on the fractionalized state only.
}
\label{QEDc2_phased_2d_hfill}
\end{figure}
The fractionalized insulator now occupies a larger area and extends 
all the way to infinitesimally small $J_{\rm bond}/u_b$ for infinitely 
large $K_{\rm ring}/u_\psi$ as shown in 
Fig.~\ref{QEDc2_phased_2d_hfill}.
It is also more stable for large $J_{\rm bond}/u_b$ due to frustration 
coming from the Berry phase terms in the corresponding Ising gauge
theory\cite{SV}.

In the model Eq.~(\ref{Hboson}), chargons are bosonic excitations.
However, as noted above, the background couplings are weakened
in the vicinity of a chargon.  It is then plausible that the
corresponding $K_{\rm ring}' < 0$ in some model, in which
case it is energetically favorable for a vison to bind to
the chargon thus forming a fermionic excitation.  This
possibility is indeed realized when we modify our model
slightly by allowing some frustration in hopping\cite{unpub}.
Thus, we obtain an explicit boson model that has an insulating phase 
with fermionic excitations carrying fractional charge.  
If this unusual insulator is doped, we would obtain a Fermi liquid, 
i.e., a metallic phase in a boson model, a true Bose metal.

\begin{figure}
\includegraphics[width=2.6in]{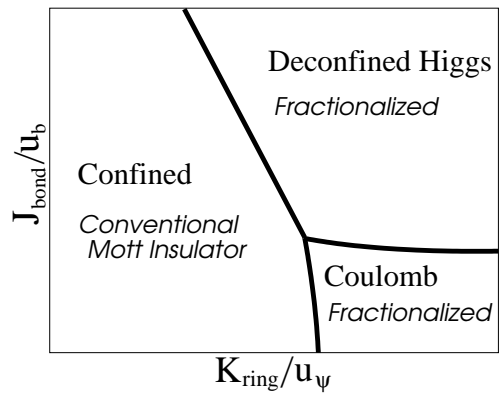}
\vskip -3mm
\caption{Same as in Fig.~\ref{QEDc2_phased_2d} but in three dimensions.
There is an additional fractionalized phase, the ``Coulomb'' phase, 
with the distinct excitations being a gapped charge $\frac{q_b}{2}$ 
chargon, gapless photon, and a gapped monopole.
}
\label{QEDc2_phased_3d}
\end{figure}

\vskip 0.1mm 
\textbf{Three dimensions:}
Consider now the 3D version of our boson model on a bond-centered 
cubic lattice.  Proceeding as before, the large $U$ Mott insulating 
states are described by the effective Hamiltonian 
$H_{\rm eff}^{(0)}$, Eq.~(\ref{Heff}),
which is equivalent to the $(3+1)$D version of the compact 
$U(1)$ gauge theory coupled to a charge 2 scalar.  
The phase diagram is shown in Fig.~\ref{QEDc2_phased_3d}, 
and now has three distinct phases.
The ``confined'' phase is the conventional Mott insulator.
The ``deconfined Higgs'' phase is similar to the fractionalized 
phase in two dimensions.  The distinct excitations here are 
gapped charge $\frac{q_b}{2}$ chargon and neutral $Z_2$ vortex loop
(vison).  The vison excitation energy is proportional to the loop
length, and the loops do not proliferate up to a finite temperature.  
Thus, in this phase in 3D, the topological order is stable for small 
finite temperature (unlike 2D where 
a finite density of thermally 
excited point visons destroys the
topological order  at any non-zero temperature). 

Finally, the ``Coulomb'' phase is also fractionalized.  
In this phase, the low-energy theory in the ground state 
sector is that of the pure gauge $(3+1)$D compact QED in its 
Coulomb phase, and has a gapless linearly dispersing gauge boson 
(photon) and a gapped topological point defect (monopole) 
as its distinct excitations.
The charged sector has charge $\frac{q_b}{2}$ excitations above 
the gap $U$, but these now interact via an {\em emergent} long-range 
Coulomb interaction.  It is quite surprising that a simple Hamiltonian 
like Eq.~(\ref{Hboson}) can have have such unusual phase.

\vskip 0.1mm
\textbf{Discussion:}
The most intriguing aspect of this paper is the simplicity
of the Hamiltonians that realize a variety of unconventional
quantum phases, and the possibility that such systems may actually 
be made in a laboratory.

In 2D one likely realization may be a Josephson 
junction array.  The particular Hubbard terms can in principle
be achieved by controlling the electrostatics of the
islands\cite{JJK,Iof}.  Studies of the superfluid-insulator 
transition can provide indirect information on the nature of the 
insulating phase.  Indeed, as the superfluid transition from the 
fractionalized phase occurs due to condensation of charge 
$\frac{q_b}{2}$ chargons, the corresponding universal conductivity 
will be $\frac{1}{4}$ that of the transition from the 
conventional insulator to the superfluid. 
Other possible techniques for detecting fractions of charge 
are discussed in Ref.~\onlinecite{toexp}.  
Of direct relevance for the implementation of topologically protected 
qubits is the vison trapping experiment:  A $2\pi$-vortex remains 
trapped in a hole in the system even when the system is cycled
from the superfluid to the fractionalized insulator and back.
For this to work, one needs to go to temperatures well below the vison 
gap.  Such flux trapping corresponds directly to the ability
of the qubit to retain its state in the experimental environment.
In this context, we want to note again that in 3D, unlike 2D,
the topological order is not destroyed by small finite temperature.

We thank P.~A.~Lee, A.~Vishwanath, and X.-G.~Wen for useful 
discussions. This work was supported by the MRSEC program of 
the National Science Foundation under grant DMR-9808941.


\vskip -5mm
\begin{thebibliography}{10}


\bibitem{RSSpN} \vskip -4mm
N.~Read and S.~Sachdev, Phys. Rev. Lett. {\bf 66}, 1773 (1991);
X.-G.~Wen, Phys. Rev. B {\bf 44}, 2664 (1991).

\bibitem{MoeSon}
R.~Moessner and S.~L.~Sondhi, Phys. Rev. Lett. {\bf 86}, 1881 (2001).

\bibitem{BalMPAFGir}
L.~Balents, M.~P.~A.~Fisher, and S.~M.~Girvin, cond-mat/0110005 
(unpublished).

\bibitem{frcmdl} 
T.~Senthil and O.~I.~Motrunich, cond-mat/0201320.

\bibitem{PWA} P.~W.~Anderson, Science {\bf 235}, 1196 (1987); 
S.~Kivelson, D.~S.~Rokhsar, and J.~Sethna, 
Phys. Rev. B {\bf 35}, 8865 (1987);
T.~Senthil and M.~P.~A.~Fisher, Phys. Rev. B {\bf 62}, 7850 (2000).

\bibitem{Kit} 
A.~Kitaev, quant-ph/9707021.

\bibitem{Iof}
L.~B.~Ioffe\etal, Nature {\bf 415}, 503 (2002).

\bibitem{Wen:light} 
X.-G.~Wen, Phys. Rev. Lett. {\bf 88}, 011602 (2002).

\bibitem{JJK} Electrostatics that favors charge neutrality in each 
cluster can be realized, e.g., as in the ``JJK'' model of 
Ref.~\onlinecite{Iof}, with which our 2D model bears close 
resemblance.  

\bibitem{FraShe}
E.~Fradkin and S.~H.~Shenker, Phys. Rev. D {\bf 19}, 3682 (1979). 

\bibitem{Wen2}
X.-G.~Wen, Int. J. Mod. Phys. B {\bf 4}, 239 (1990).

\bibitem{topth}
T.~Senthil and M.~P.~A.~Fisher, Phys. Rev. B {\bf 63}, 134521 (2001).

\bibitem{toGutzw} 
D.~A.~Ivanov and T.~Senthil, cond-mat/0204043.

\bibitem{unpub} 
O.~I.~Motrunich and T.~Senthil (unpublished).

\bibitem{SV}
S.~Sachdev and M.~Vojta, J. Phys. Soc. Jpn. {\bf 69}, Supp. B, 1 (2000).

\bibitem{toexp}
T.~Senthil and M.~P.~A.~Fisher, Phys. Rev. Lett., {\bf 86}, 292 (2001);
Phys. Rev. B {\bf 64}, 214511 (2001).

 
\end{thebibliography}
\end{document}